\documentclass[aps,pra,twocolumn]{revtex4-1}

\usepackage[english]{babel}
\usepackage[T1]{fontenc}
\usepackage[utf8]{inputenc}

\usepackage{bbm}  
\usepackage{graphicx}
\usepackage{amsmath}
\usepackage{amssymb}
\usepackage{amsfonts}
\usepackage{braket}
\usepackage{booktabs}
\usepackage{multirow}
\usepackage{simplewick}
\usepackage{bm} 
\usepackage{xcolor}
\usepackage{ulem}
\usepackage{threeparttable}
\usepackage{subfig}
\usepackage{float}

\usepackage{cellspace}
\usepackage[autostyle=true]{csquotes}

\graphicspath{{figures/}}

\def\Emin{\hat{\mathbb{E}}_{\text{min}}}
\def\Emax{\hat{\mathbb{E}}_{\text{max}}}
\def\ELow{\hat{\mathbb{E}}_{\text{L\"ow}}}

\begin{document}                                                                                                                                                                   
                                                                                                                                                                                   
\title{Analysis of Different Sets of Spin-Adapted Substitution Operators\\ in Open-Shell Coupled Cluster Theory}
\author{Nils Herrmann}
\email{N.Herrmann@uni-koeln.de}
\author{Michael Hanrath}
\email{Michael.Hanrath@uni-koeln.de}

\affiliation{%
Institute for Theoretical Chemistry, University of Cologne, 
Greinstra\ss e 4, 50939 Cologne, Germany
}%

\date{\today}

\begin{abstract}
In spin-adapted open-shell coupled cluster (CC) theory, the choice of spin-free spatial substitution operators is generally not unique. Due to an increasing linear dependence of the cluster operator (with increasing substitution level), the options to span identical linear spaces increase rapidly. 
In this work several sets of non-orthogonal as well as orthogonal spin-adapted substitution operators are generated and used in consecutive Configuration Interaction (CI) and CC calculations. All (full) operator sets were generated to span the same linear space. The results are analyzed in terms of the produced wave function quality and the amount of recovered correlation energy w.r.t. full CI.
In particular, the influence of different amounts of spectators, the influence of orthogonality as well as the effect of spin incompleteness was investigated. It was found that CC calculations involving fewer spectators lead to more accurate results in general. Here correlation energy differences of up to 0.32\% for minimal to maximal spectating sets were obtained. As expected, all conducted calculations led to identical results for non-orthogonal and orthogonal operator sets. Spin completeness on the other hand was found to be of great importance. Spin-incomplete Cluster operators led to significant errors in both the correlation energies and the FCI overlap.   
\end{abstract}

\maketitle

\section{Introduction}

Since its development in the late 50s\cite{Coester1958, Coester1958a}, coupled cluster (CC) has emerged as one of the most accurate quantum chemical methods with applications in all aspects of modern quantum chemistry.

For more than 25 years it is well understood that spin contamination appears in open-shell CC wave functions (even for pure-spin references, e.g. ROHF)\cite{Stanton1994, Krylov2000, Li2000} and leads to a broken spin symmetry. Ever since, research efforts were applied to solve the latter problem.\cite{Mukherjee1975, Rittby1988, Yuan2000, Neogrady1992, Neogrady1994, Neogrady1995, Szalay1997, Szalay2000, Berente2002, Heckert2006, Wilke2011, Knowles1993, Knowles2000, Nakatsuji1978, Janssen1991, Li1993, Li1994, Piecuch1994, Li1994a, Li1995, Li1995a, Li1995b, Jeziorski1995, Jankowski1999, Nooijen1996, Nooijen2001, Sen2012, Datta2008, Datta2009, Datta2011, Datta2013, Datta2014, Datta2015, Datta2019, Herrmann2020a, Krylov2017}. Several advancements were also made in the special case of a closed-shell reference.\cite{Paldus1977, Adams1979, Ghosh1982, Aadnan1982, Takahashi1985, Jeziorski1988, Piecuch1989, Piecuch1990, Piecuch1992, Piecuch1995, Kondo1995, Kondo1996, Matthews2013, Matthews2015} This work, however, shall focus on the more general open-shell case, only.

Standard spinorbital implementations of open-shell CC are readily available in the literature (e.g. Mukherjee et al.\cite{Mukherjee1975}) at the cost of spin contamination. To remove the latter, spin-projected\cite{Rittby1988, Yuan2000} and spin-restricted\cite{Neogrady1992, Neogrady1994, Neogrady1995, Szalay1997, Szalay2000, Berente2002, Heckert2006, Wilke2011} methods were developed. Neogr\'{a}dy et al. derived spin-adapted variants of linear CCSD\cite{Neogrady1992}, full CCSD\cite{Neogrady1994} and CCSD(T) employing a perturbative triples correction\cite{Neogrady1995} for the doublet spin state while a spin-restricted augmentation of the standard spinorbital approach was developed by Szalay and Gau\ss{}.\cite{Szalay1997} In their approach, spin constraints, which guarantee a correct $\hat{S}^2$ expectation value, are forced on the CC wave function. To make the approach computationally feasible, the latter constraints are only followed in a truncated subspace such that the approach is not rigorously spin-adapted. Later, a comparison between spin-restricted and rigorously spin-adapted approaches was given\cite{Heckert2006}, the approach was augmented to treat excited states\cite{Szalay2000}, include full triples\cite{Berente2002} and has been used in an explicitly correlated R12 framework.\cite{Wilke2011} 

Furthermore, Knowles and Werner developed partial spin-adapted CCSD variants\cite{Knowles1993, Knowles2000}, where a mix of spinorbital and spatial substitution operators was used such that the linear CC terms were rigorously spin-adapted. The non-linear terms, however, were not. 

In contrast to spinorbital approaches, the idea of a symmetry-adapted or spin-adapted cluster operator $\hat{T}$, originally suggested by Nakatsuji and Hirao\cite{Nakatsuji1978}, is to define a cluster operator, which commutes with both the total spin operator $\hat{S}^2$ and the spin projection $\hat{S}_z$ with
\begin{equation}
\left[\hat{T}, \hat{S}^2\right] = \left[\hat{T}, \hat{S}_z\right] = 0\,,
\end{equation}
such that spin quantum numbers $S$ and $S_z$ are automatically conserved. Published approaches may be sorted into the following groups:
\begin{itemize}
\item[(i)] plain non-orthogonal approaches\cite{Janssen1991} employing solitary spatial substitutions $\hat{E}$, which lead to non-orthogonal CSFs,
\item[(ii)] orthogonal approaches\cite{Li1993, Li1994, Piecuch1994, Li1994a, Li1995, Li1995a, Li1995b, Jeziorski1995, Jankowski1999} using machinery of unitary group theory to derive linear combinations of spatial substitutions $\hat{E}$, which generate orthogonal CSFs,
\item[(iii)] explicitly normal-ordered approaches\cite{Nooijen1996,Nooijen2001, Sen2012}, where the wave operator $e^{\hat{T}}$ is assumed to be normal-ordered, as originally suggested in the late 70s by Lindgren\cite{Lindgren1978} and 
\item[(iv)] explicitly normal-ordered approaches reintroducing contractions of spectating substitutions via a combinatoric cluster expansion\cite{Datta2008, Datta2009, Datta2011, Datta2013}, which has recently been augmented to determine first order properties and hyperfine coupling constants\cite{Datta2014, Datta2015, Datta2019}.
\end{itemize}

All of the aforementioned implementations seem to be limited to the doubles truncation 
and $S \le 1$ (triplet spin state). This is probably due to 
\begin{itemize} 
\item[(1)] the need for spectating substitutions to reach spin completeness, which complicates the CC theory itself (non-vanishing $\contraction{}{\hat{T}}{}{\hat{T}}\hat{T}\hat{T}$ and $\contraction{}{\hat{T}}{}{\mathcal{\hat{H}}}\hat{T}\mathcal{\hat{H}}$ contractions lead to a non-terminating BCH series) and 
\item[(2)] the need to reformulate the cluster operator for every spin quantum number $S$ as well as its increasing linear dependence with increasing substitution rank.  
\end{itemize}

In our previous work\cite{Herrmann2020a}, we proposed a solution to (2) based on L\"owdin's\cite{Lowdin1955, Lowdin1964} projection operator method of spin eigenfunction generation. We developed a canonical operator generation scheme to readily build non-orthogonal cluster operators of arbitrary order, which are applicable to arbitrary (high spin) states. The generated cluster operators, denoted L\"owdin-based operators, however, are in no way unique. Due to the increasing linear dependency of the spatial substitution operators with increasing substitution rank, very different sets of same-space-spanning cluster operators are obtainable. Another source of non-uniqueness is an optional subsequent orthogonalization of the obtained operator.

It is the intent of this work to systematically investigate the effects of spin-adapted cluster operator differences on the wave function quality and the amount of recovered correlation within the non-orthogonal as well as the orthogonal spin-adapted open-shell CC framework. Special attention will be given to our automatically generated L\"owdin-based cluster operators.\cite{Herrmann2020a}

\section{Methodology}

In this work, several differing sets of spatial substitution operators are applied in CI and CC calculations. The results are analysed with respect to (i) the wave function quality through means of the overlap to the FCI wave function and (ii) the amount of recovered correlation energy compared to FCI. The analysed operator sets are composed of different amounts of spectators as well as orthogonal or non-orthogonal operators. 
In the following subsections we discuss the generation
of various different spin-adapted operator sets. 

\subsection{Generating Operator Sets}
\label{GeneratingOperatorSets}

Throughout this paper we will denote doubly occupied spatial orbitals as $i,j,k\ldots\in\mathbb{O}$, singly occupied spatial orbitals as $v,w,x\ldots\in\mathbb{A}$ and empty virtual spatial orbitals as $a,b,c\ldots\in\mathbb{V}$. Arbitrary spatial substitutions $\hat{E}$ may incorporate substitutions from the joined spaces $p_1,p_2,p_3\ldots\in\mathbb{O}\cup\mathbb{A}$ to the joined spaces $q_1,q_2,q_3\ldots\in\mathbb{A}\cup\mathbb{V}$. They may be defined by the spin integration of spinorbital substitution operators $\hat{X}$ via  
\begin{align}
\label{EDefinition.eq}
\hat{E}_{p_1\ldots p_{\nu}}^{q_1\ldots q_{\nu}} &= \sum_{\sigma_1 = \alpha , \beta}\ldots\sum_{\sigma_{\nu} = \alpha , \beta} \hat{X}_{p_1\sigma_1 \ldots p_{\nu}\sigma_{\nu}}^{q_1\sigma_1 \ldots q_{\nu}\sigma_{\nu}} \\
\notag
&= \sum_{\sigma_1 = \alpha, \beta} \hat{a}_{q_1\sigma_1}^{\dagger} \left( \ldots \left( \sum_{\sigma_{\nu} = \alpha, \beta} \hat{a}_{q_{\nu}\sigma_{\nu}}^{\dagger} \hat{a}_{p_{\nu}\sigma_{\nu}} \right) \ldots \right) \\
&\phantom{==}\hat{a}_{p_1\sigma_1}\,.
\notag
\end{align}

Since (\ref{EDefinition.eq}) employs pairwise spin summations only, it is 
\begin{equation}
\hat{E}_{p_1\ldots p_{\nu}}^{q_1\ldots q_{\nu}} = \hat{E}_{\hat{P}(p_1\ldots p_{\nu})}^{\hat{Q}(q_1\ldots q_{\nu})}\quad\forall_{\hat{P} = \hat{Q} \in \mathbb{S_{\nu}}}\,,
\end{equation}
where $\hat{P}$ and $\hat{Q}$ denote arbitrary index permutations
of the symmetric group $\mathbb{S_{\nu}}$. Therefore, all non trivially identical spatial substitution operators of rank $\nu$ are e.g. given by
\begin{equation}
\bigcup_{\hat{P}\in\mathbb{S}_{\nu}} \hat{E}_{p_1 \le \ldots \le p_{\nu}}^{\hat{P}(q_1\ldots q_{\nu})}\,,
\end{equation}
where all annihilators are sorted in ascending order. 

In this work a total of three different (but complete) operator sets employing solitary spatial substitutions $\hat{E}$ were generated. For the first two sets, we generated all substitutions $\hat{E}_{p_1 \le \ldots \le p_{\nu}}^{\hat{P}(q_1\ldots q_{\nu})}$ for all $\hat{P}\in\mathbb{S}_{\nu}$ and for a $\nu$ of 1 up to the number of electrons $n$. These substitutions were then sorted into subsets of configuration generating operators 
\begin{equation}
\hat {\mathbb E}_\lambda = \left\{ \hat E_{\ldots}^{\ldots} \; | \; 
\Lambda\left(\hat E_{\ldots}^{\ldots} \ket{\Psi_0}\right) = \ket{\lambda} \right\}
\end{equation}
with $\Lambda(\ket{\Phi})$ extracting the spatial occupation (configuration) of determinant $\ket{\Phi}$ and $\ket{\Psi_0}$ the reference
CSF. These subsets are usually much larger than the required spin degeneracy of $f(O,S)$ for the given spatial configuration (see e.g.\cite{Pauncz1979}) with 
\begin{equation}
f(O,S) = \binom{O}{\frac{O}{2} - S} - \binom{O}{\frac{O}{2} - S - 1}
\end{equation}
for spin quantum number $S$ and $O$ open shells. This way we may choose different operators from the subsets to constitute our cluster operators while spanning the same space. In order to build two operator sets, which are as different as possible, we sort the individual $\hat {\mathbb E}_\lambda$ with respect to operator rank
ascending and descending.\footnote{Please note that rank here implies the nominal tensor rank and not the order of spatial substitution since the latter would be identical for all operators in $\hat {\mathbb E}_\lambda$.} From the sorted subsets we may then choose (i) the first $f(O,S)$ linearly independent operators such that we arrive at a final set incorporating a minimal amount of spectating indices -- denoted by $\Emin$ -- or (ii) the last $f(O,S)$ linearly independent operators such that we arrive at a final set incorporating a maximal amount of spectators -- denoted by $\Emax$.

The third operator set is composed of L\"owdin-based operators, which were generated using an automated approach we explicitly derived in our previous work.\cite{Herrmann2020a} In this approach, a set of distinct $\hat{E}$ operator prototypes is permuted and/or augmented utilizing L\"owdin's projection operator method for the generation of non-orthogonal spin eigenfunctions\cite{Lowdin1955, Lowdin1964} to arrive at a linear independent and spin-complete cluster operator. In the following we will refer to this operator set as $\ELow$. 

\subsection{(Non-)Orthogonality}
\label{Orthogonality}

Since the generated operator sets are composed of single spatial substitutions $\hat{E}$, they produce non-orthogonal CSFs in general. 
In the literature, several ans\"atze utilizing orthogonal linear combinations of $\hat{E}$ substitution operators are known (see e.g. \cite{Li1993, Jeziorski1995}). 
In terms of complexity and efficiency, 
non-orthogonal cluster operators are more convenient than orthogonalized cluster operators\cite{Matthews2013, Springer2019} as may be easily recognized by their simpler form.

For a linear wavefunction ansatz like e.g. CI the wavefunction and energy are directly
governed by the spanned function space. For a non-linear wavefunction approach like e.g. CC
this relation might a priori not be completely obvious. To this end we take a closer look and
consider a non-orthogonal spin-adapted cluster operator $\hat{T}$ with 
\begin{equation}
\label{Tfromvecs.eq}
\hat{T} = \mathbf{t}^{\top} \mathbf{E} = \sum_i t_i \hat{E}_i\, ,
\end{equation}
where $\mathbf{t}$ and $\mathbf{E}$ shall denote vectors composed of the amplitudes $t_i$ 
and the non-orthognal spatial substitution operators $\hat{E}_i$, respectively. 
The same space spanning orthogonal operators $\mathbf{E'}$ may be assembled by
\begin{equation}
\mathbf{E'} = \uuline{\mathbf{X}}\mathbf{E}\,.
\end{equation}
Assuming the $\{\hat{E_i}\}$ to be linearly independent also
the inverse transformation $\uuline{\mathbf{X}}^{-1}$ exists.
This allows for the insertion of the identity $\uuline{\mathbf{1}} = \uuline{\mathbf{X}}^{-1} \uuline{\mathbf{X}}$ in (\ref{Tfromvecs.eq}) 
and leads to 
\begin{equation}
\hat{T} = \underbrace{\mathbf{t}^{\top} \uuline{\mathbf{X}}^{-1}}_{\mathbf{t'}^{\top}} \underbrace{\uuline{\mathbf{X}} \mathbf{E}}_{\mathbf{E'}} = \mathbf{t'}^{\top} \mathbf{E'} = \hat{T}'\,,
\end{equation}
such that the change of operator basis $\mathbf{E}\rightarrow\mathbf{E}'$ is accompanied by a change of its representation $\mathbf{t}\rightarrow\mathbf{t}'$. For the same arguments the powers of $\hat T$ and $\hat T'$ remain identical.
Therefore, any CC calculation (if properly converged) involving either $\hat{T}$ or $\hat{T}'$ should lead to identical wave functions.

In case of the three operator sets $\ELow$, $\Emin$ and $\Emax$ (c.f. subsection \ref{GeneratingOperatorSets}), identical results are expected for their orthogonalized counterparts $\ELow'$, $\Emin'$ and $\Emax'$, respectively. However, it is important to note that different results are (as in the non-orthogonal case) expected for a direct comparison of $\ELow'$, $\Emin'$ and $\Emax'$ since they contain different substitutions which can not be transformed into one another by a linear transformation.
To summarize it is $\hat{\mathbb{E}}_M \sim \hat{\mathbb{E}}'_M$
while $\hat{\mathbb{E}}_M \not\sim \hat{\mathbb{E}}_{N}$ and
$\hat{\mathbb{E}}'_M \not\sim \hat{\mathbb{E}}'_{N}$  for $M\neq N$.

To check these equalities (and inequalities), all generated operator sets were explicitly orthonormalized using an appropriately chosen transformation $\uuline{\mathbf{X}}$, which transforms to an orthonormal CSF basis $\uuline{\mathbf{X}}\mathbf{E}\ket{\Psi_0}$. This basis was constructed by merging spatial configurations with the appropriate number of spin eigenfunctions generated via a genealogical coupling algorithm (see e.g. \cite{Pauncz1979}). CC calculations employing both, the non-orthogonal as well as the orthogonalized cluster operator were then compared as outlined in the following subsections.  

\subsection{CI and CC in FCI Representation}
\label{FCIRepresentation}

In a general complete non-orthogonal basis set $\{\ket{i}\}$, arbitrary operators $\hat{O}$ are given by 
\begin{equation}
\hat{O} = \sum_{ij} \ket{i} \left( \uuline{\mathbf{S}}^{-1}\cdot\uuline{\mathbf{O}}\cdot\uuline{\mathbf{S}}^{-1} \right)_{ij} \bra{j}\,,
\end{equation}
where $\uuline{\mathbf{S}}^{-1}$ denotes the inverse overlap matrix $\uuline{\mathbf{S}}$ while the elements of $\uuline{\mathbf{O}}$ are given by 
\begin{equation}
\left(\uuline{\mathbf{O}}\right)_{ij} = \braket{i|\hat{O}j}\,.
\end{equation}

Therefore, representation of general operator products $\hat{O}_1\cdot\hat{O}_2\cdot \ldots \cdot \hat{O}_m$ are given by 
\begin{equation}
\label{NonOrthogonalOperatorProduct.eq}
\braket{i|\hat{O}_1\hat{O_2}\ldots\hat{O}_m j} = \left(\uuline{\mathbf{O}_1}\cdot\uuline{\mathbf{S}}^{-1} \cdot \uuline{\mathbf{O}_2}\cdot \uuline{\mathbf{S}}^{-1}  \ldots  \uuline{\mathbf{S}}^{-1} \uuline{\mathbf{O}_m}\right)_{ij}\,.
\end{equation}

Please note that an orthonormal CSF basis leaves $\uuline{\mathbf{S}} = \uuline{\mathbf{S}}^{-1} = \uuline{\mathbf{1}}$ such that (\ref{NonOrthogonalOperatorProduct.eq}) becomes the much more familiar form of matrix representation products $\uuline{\mathbf{O}_1}\cdot\uuline{\mathbf{O}_2}\cdot\ldots\cdot\uuline{\mathbf{O}_m}$, only. 

In general, the CI and CC wave functions are given by
\begin{align}
\ket{\Psi_{\text{CI}}} &= \left(1 + \hat{C}\right) \ket{\Psi_0}
\\
\ket{\Psi_{\text{CC}}} &= e^{\hat{T}}\ket{\Psi_0}\,,
\end{align}
respectively,
where $\ket{\Psi_0}$ denotes the reference CSF as well as
 $\hat{C}$ and $\hat{T}$ the CI and cluster operators, respectively.
The latter are composed of the particular $\hat{E}$ operators from subsection \ref{GeneratingOperatorSets}. 

All calculations (CI and CC) analyzed in this paper were conducted in the complete FCI CSF basis such that the cluster operator $\hat{T}$, the CI operator $\hat{C}$, the Hamiltonian $\hat{\mathcal{H}}$ as well as the identity $1$ were represented in this basis as matrix representations $\uuline{\mathbf{T}}$, $\uuline{\mathbf{C}}$, $\uuline{\mathbf{H}}$ and $\uuline{\mathbf{S}}$, respectively.

Represented in a general non-orthogonal basis, the CI equations take the form 
\begin{equation}
\label{CI.eq}
\left(\uuline{\mathbf{H}} + \uuline{\mathbf{H}}\cdot \uuline{\mathbf{S}}^{-1} \cdot \uuline{\mathbf{C}}\right)_{X,0} - E_{\text{CI}} \left(\uuline{\mathbf{S}} + \uuline{\mathbf{C}}\right)_{X,0} = 0\,,
\end{equation}

with $X \in \mathbb{N}_0$, $X \le |\textrm{ls} ( \{\hat E_i\} \ket{\Psi_0} )|$.
The reference CSF $\ket{\Psi_0}$ associated with the energy projection
is chosen to be the first basis function $X=0$. Later basis functions $X>0$ correspond
to the residual projections.

Analogously, the CC equations are given by the BCH series now involving matrix commutators only with
\begin{equation}
\label{CC.eq}
\left(
\uuline{\mathbf{H}} 
+ \uuline{\mathbf{H}}\cdot\uuline{\mathbf{S}}^{-1}\cdot\uuline{\mathbf{T}}
- \uuline{\mathbf{T}}\cdot\uuline{\mathbf{S}}^{-1}\cdot\uuline{\mathbf{H}}
+ \ldots
\right)_{X,0} = \delta_{X,0} E_{\text{CC}}
\end{equation} 
In this work, the linear (CI) as well as non-linear (CC) equation systems (\ref{CI.eq}) and (\ref{CC.eq}) were solved using the GNU Scientific Library\cite{GSL} and its \texttt{gsl\_multiroot\_fsolver\_hybrids} module.  

\subsection{Analytics: Spin Projection Error}
\label{SpinProjectionError}

To check if the used operator sets are producing true spin eigenfunctions, an error estimate $\epsilon$ with 
\begin{equation}
\label{SpinProjectionError.eq}
\epsilon = \sqrt{1 - \Braket{\Psi_{\text{CI/CC}}^{S,S_z}|\Psi_{\text{CI/CC}}^{S,Sz}}}
\end{equation}
analogously to \cite{Hanrath2009} was calculated. Here, $\Ket{\Psi_{\text{CI/CC}}^{S,Sz}}$ denotes the projected (after normalization $\braket{\Psi_{\text{CI/CC}} | \Psi_{\text{CI/CC}}} = 1$) CI or CC wave function with 
\begin{equation}
\Ket{\Psi_{\text{CI/CC}}^{S,Sz}} = \hat{\mathcal{P}}_{\text{CSF}}^{S,S_z}\ket{\Psi_{\text{CI/CC}}}\,,
\end{equation}
where $\hat{\mathcal{P}}_{\text{CSF}}^{S,S_z}$ shall denote a projector onto the full CSF basis of the desired quantum numbers $S$ and $S_z$. As stated in \cite{Hanrath2009}, the spin projection error $\epsilon$ is a much stronger check for spin state purity then e.g. the deviation of the $\hat{S}^2$ expectation value since the latter might benefit from error cancellations. 

\subsection{Analytics: FCI Overlaps}
\label{FCIOverlap}

To investigate the influence of different cluster and CI operators on the overall quality of the CI and CC wave functions, the overlap of the latter wave functions to the exact FCI wave function was calculated. Therefore, both wave functions were normalized. In the untruncated CI and CC limits, the wave functions should be identical such that the (absolute) overlap should be exactly 1. 

For cases with degenerate FCI ground state solutions, it may not be enough to consider the overlap to only one of the solutions due to arbitrary rotations in the eigenspace. Therefore, following
appendix \ref{appendix}, the overlap was maximized considering all degenerate FCI solutions using the method of Lagrange multipliers.

\section{Results}

In this section, the results of the conducted calculations are summarized and discussed. In particular, FCI overlaps $\braket{\tilde{\Psi}|\Psi_{\text{FCI}}}$ are visualized using pW values with 
\begin{equation}
\label{pW.eq}
\text{pW} = -\log_{10} \left(1 - |\braket{\tilde{\Psi}|\Psi_{\text{FCI}}}|\right)\,,
\end{equation}
which approaches infinity for $\ket{\tilde{\Psi}} = \ket{\Psi_{\text{FCI}}}$. Similarly, the amount of recovered correlation energy is visualized by pE values given by 
\begin{equation}
\label{pE.eq}
\text{pE} = -\log_{10}\left|1 - \frac{\tilde{E}_{\text{corr}}}{E_{\text{corr}}^{\text{FCI}}}\right|\,,
\end{equation} 
such that a pE value of $2$ represents $99\%$, a value of $3$ $99.9\%$, etc. recovered correlation energy.

A series of CI and CC calculations was carried out on top of converged ROHF calculations using the PySCF program package\cite{Sun2018, Sun2015} 
to analyze the wave function quality as well as the amount of recovered correlation energy comparing
\begin{itemize}
\item[(A)] the different but same-space-spanning 
non-orthogonal operator sets $\hat{\mathbb{E}}_{M}$
\item[(B)] the non-orthogonal ($\hat{\mathbb{E}}_{M}$) and orthogonal ($\hat{\mathbb{E}}'_{M}$) operator sets and 
\item[(C)] a potentially spin complete,  non-orthogonal, spin-adapted operator sets ($\hat{\mathbb{E}}_{M}$) 
\end{itemize}
for all generated operator sets $M \in \{$min, max, L\"ow$\}$ (c.f. subsection \ref{GeneratingOperatorSets}).
Detailed discussions of (A) -- (C) are given in the corresponding subsections \ref{(A)}, \ref{(B)} and \ref{(C)}. 

Due to the overlapping creator and annihilator space of the employed spatial substitutions, the BCH series is not naturally vanishing for quintuply and higher nested commutators. Therefore, all conducted CC calculations were explicitly checked for their correlation energy convergence with increasing BCH series truncation. This truncation was applied such that the correlation energy was consistent within the applied residual mean square threshold. For all calculations except $^3$P carbon and $^4$S nitrogen, a BCH series 
truncation at quadruply nested commutators was sufficient.     

\subsection{Spin-complete non-orthogonal Operator Sets: Spectator Influence}
\label{(A)}
To analyze the influence of the different but same-space-spanning sets of spatial substitution operators $\ELow$, $\Emin$ and $\Emax$ (c.f. subsection \ref{GeneratingOperatorSets}) on the wave function quality as well as the amount of recovered correlation energy, CI and CC calculations of all possible high spin states of the atomic systems Li, Be, B, C and N in the basis set 6-31G\cite{Hehre1972} were conducted. Results for all of those systems can be found in the supporting information. 

\onecolumngrid
\begin{center}
\begin{figure}[H]
\centering 
\subfloat[\label{C_3_2_CI.fig}$^3$P carbon (CI)]{\includegraphics[scale=.95]{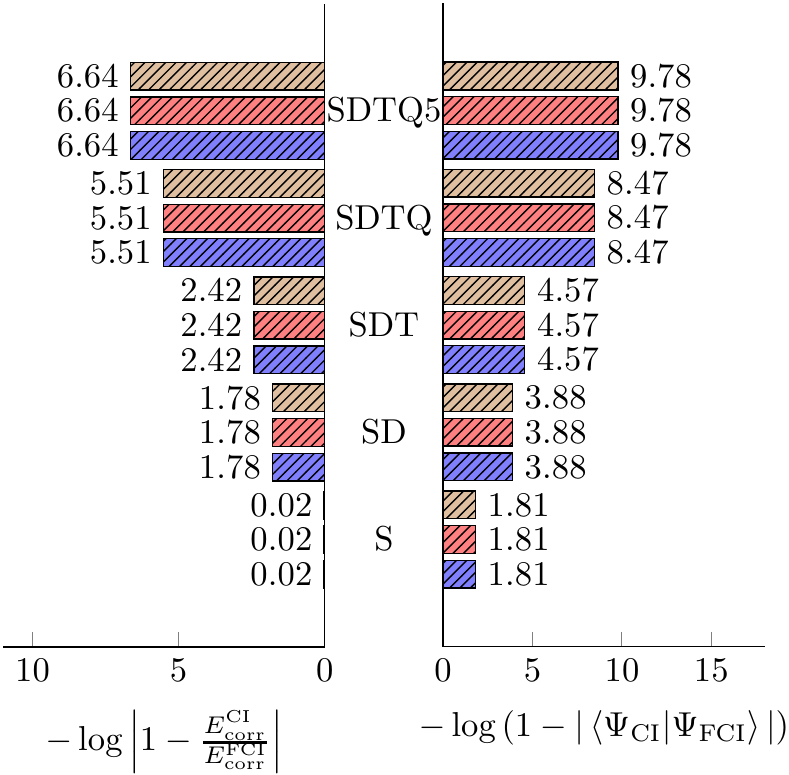}}
\quad
\subfloat[\label{C_3_2_CC.fig}$^3$P carbon (CC)]{\includegraphics[scale=.95]{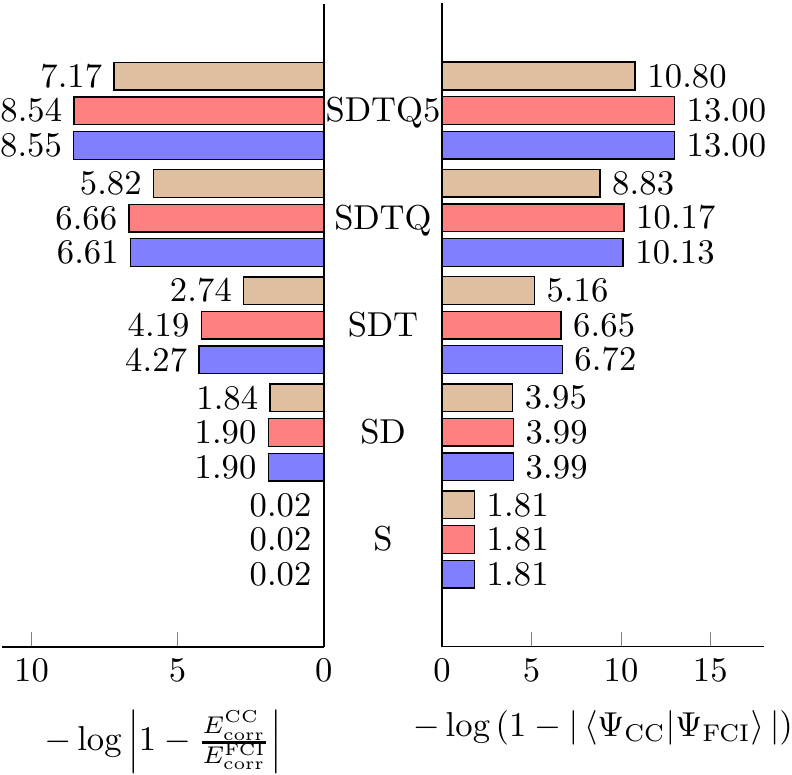}}
\\
\subfloat[\label{C_5_4_CI.fig}$^5$S carbon (CI)]{\includegraphics[scale=.95]{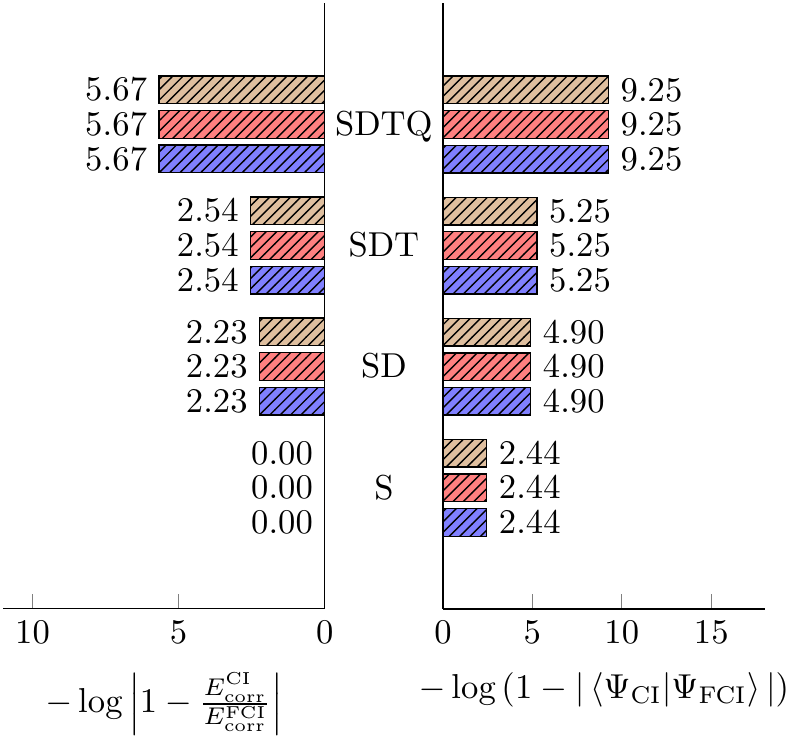}}
\quad
\subfloat[\label{C_5_4_CC.fig}$^5$S carbon (CC)]{\includegraphics[scale=.95]{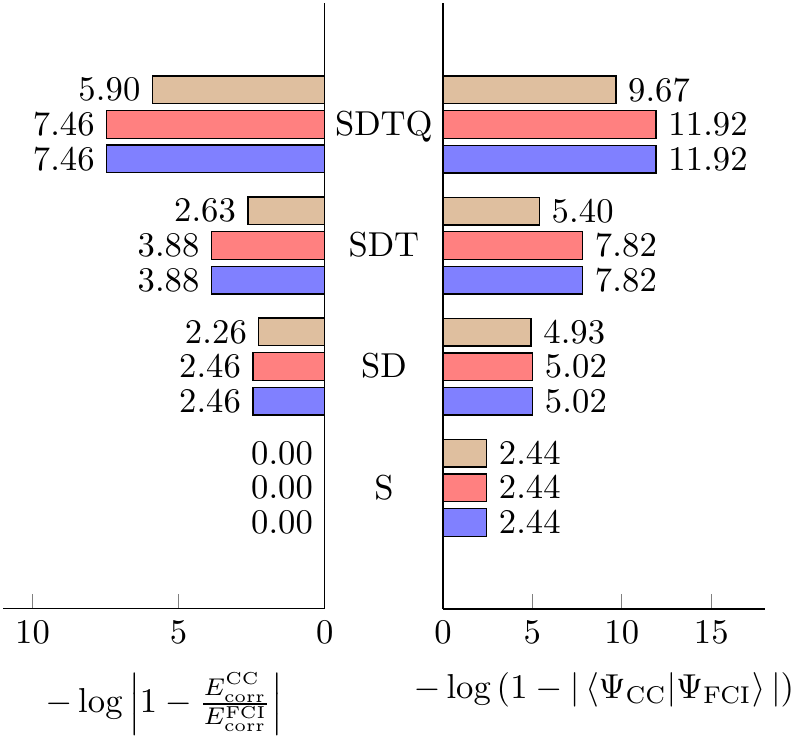}}
\\
\subfloat[\label{C_7_6_CI.fig}$^7$P carbon (CI)]{\includegraphics[scale=.95]{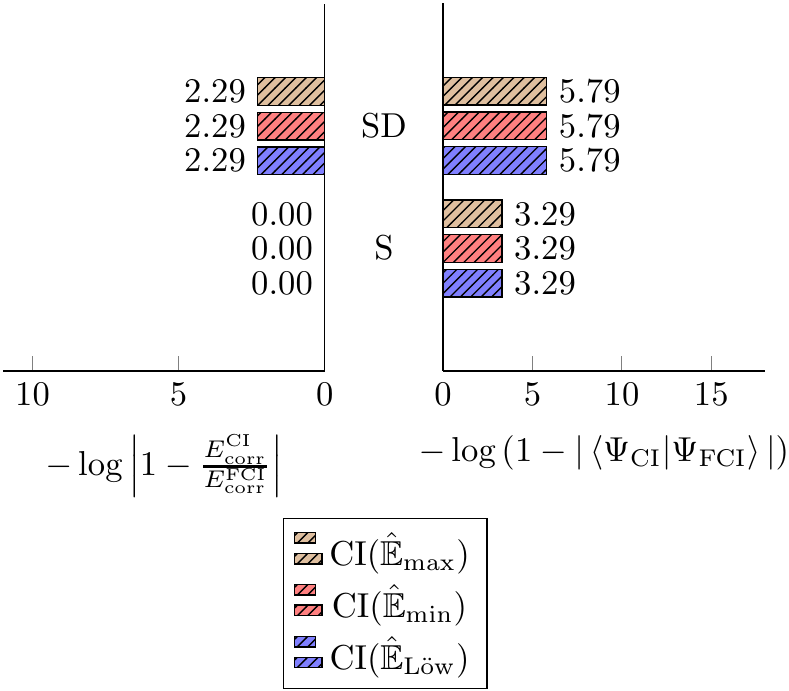}}
\quad
\subfloat[\label{C_7_6_CC.fig}$^7$P carbon (CC)]{\includegraphics[scale=.95]{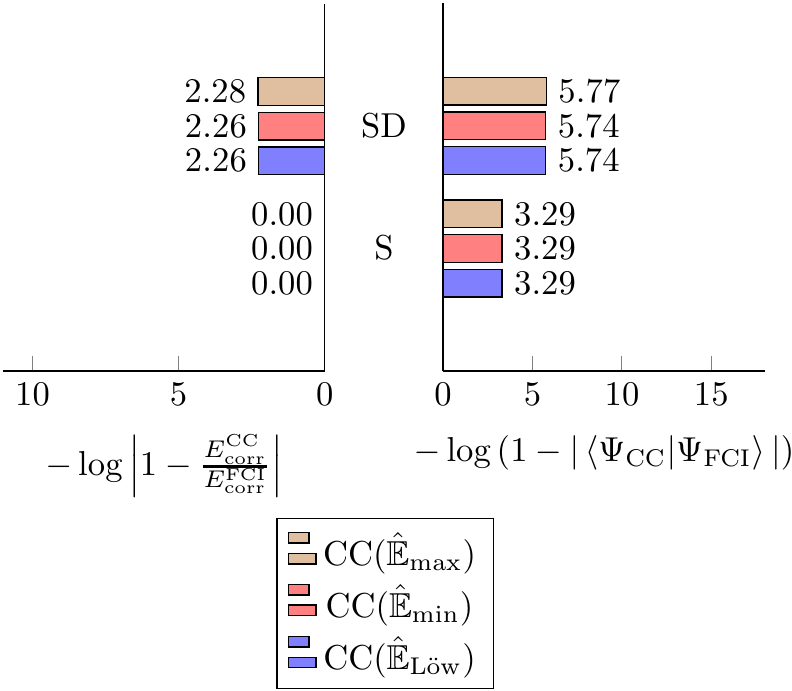}}
\caption{\label{C_I.fig}Summarized pE and pW values for spin-adapted and spin-complete CI (left) and CC (right) calculations for atomic $^3$P, $^5$S and $^7$P carbon in the 6-31G\cite{Hehre1972} basis set.}
\end{figure}
\end{center}
\begin{center}
\begin{figure}[H]
\centering 
\subfloat[\label{N_4_3_CI.fig}$^4$S nitrogen (CI)]{\includegraphics[scale=.95]{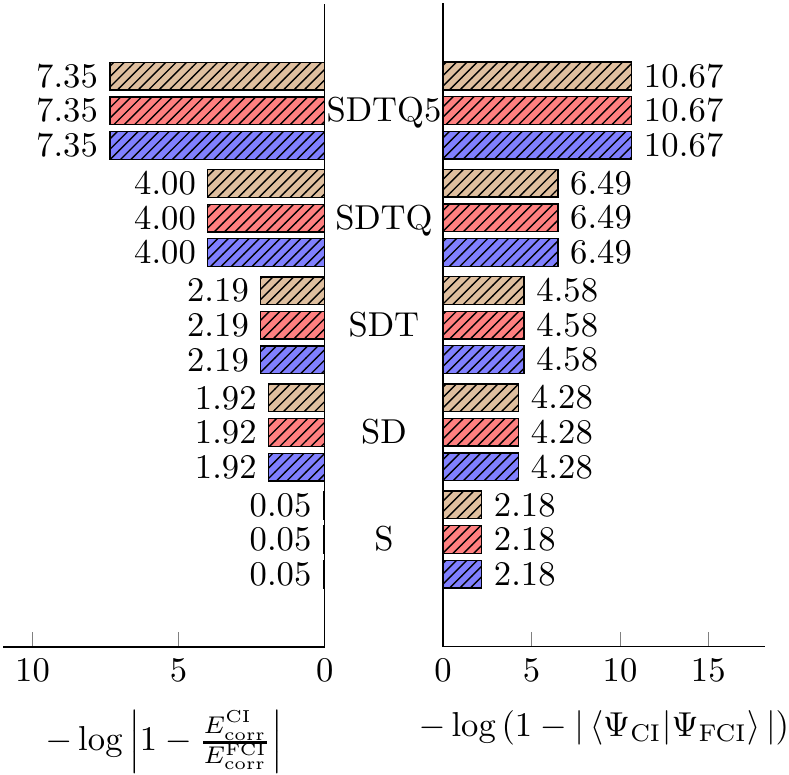}}
\quad
\subfloat[\label{N_4_3_CC.fig}$^4$S nitrogen (CC)]{\includegraphics[scale=.95]{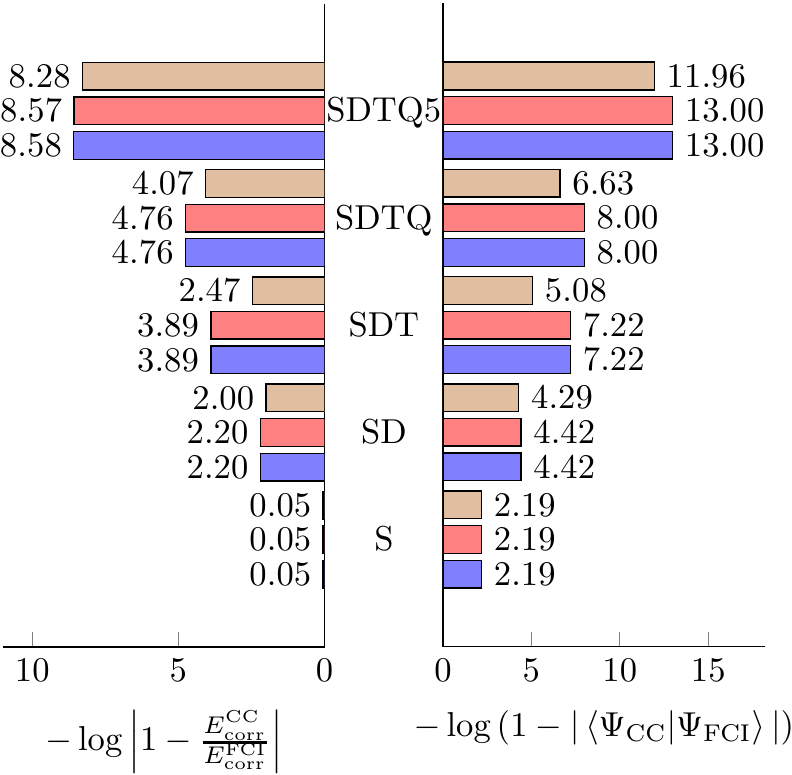}}
\\
\subfloat[\label{N_6_5_CI.fig}$^6$P nitrogen (CI)]{\includegraphics[scale=.95]{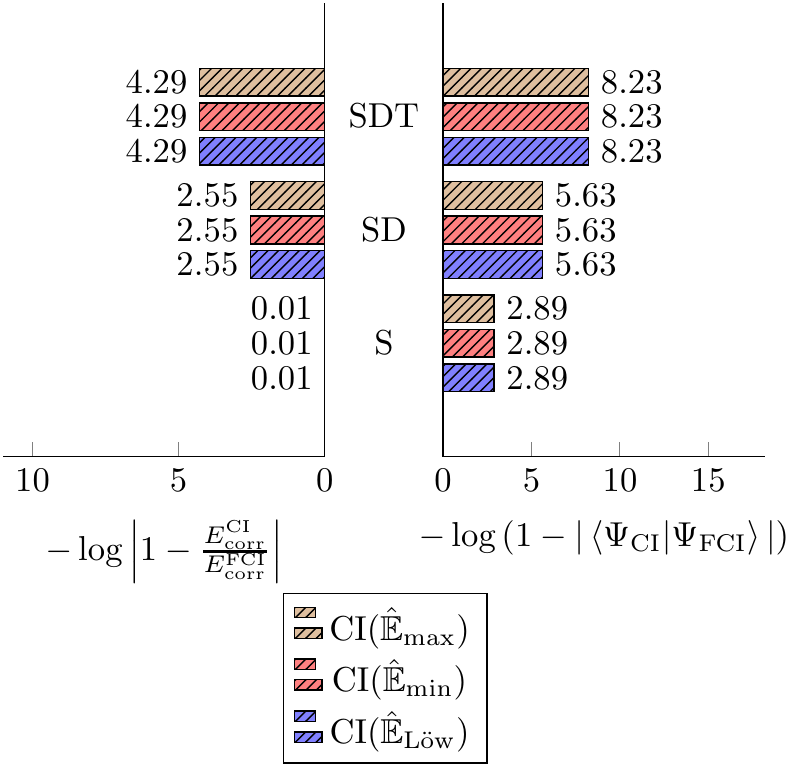}}
\quad
\subfloat[\label{N_6_5_CC.fig}$^6$P nitrogen (CC)]{\includegraphics[scale=.95]{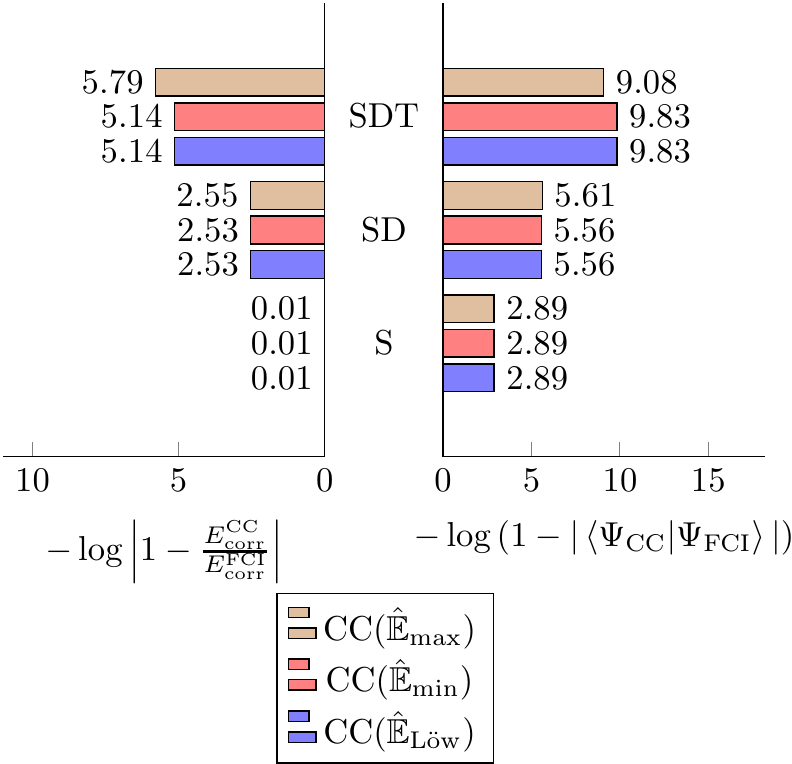}}
\caption{\label{N_I.fig}Summarized pE and pW values for spin-adapted and spin-complete CI (left) and CC (right) calculations for atomic $^4$S and $^6$P nitrogen in the 6-31G\cite{Hehre1972} basis set.}
\end{figure}
\end{center}
\twocolumngrid

For the sake of this subsection, we shall focus on the two largest systems (C and N), only. In Figures \ref{C_I.fig} and \ref{N_I.fig}, the pE and pW values for $^3$P, $^5$S and $^7$P atomic carbon
as well as $^4$S and $^6$P nitrogen are displayed, respectively. Please note that states $^7$P and $^6$P were found here (instead of the expected $^7$S and $^6$S states) due to the rather small one particle basis 6-31G. In all subfigures \ref{C_3_2_CI.fig} -- \ref{C_7_6_CC.fig} and \ref{N_4_3_CI.fig} -- \ref{N_6_5_CC.fig}, two horizontal bar diagrams show the magnitude of the pW (right bar diagram) and the pE values (left bar diagram). Furthermore, subfigures on the left-hand side display CI results (indicated by the striped pattern) while subfigures on the right-hand side display CC results.

The results for the spin-adapted and spin-complete operator sets $\ELow$, $\Emin$ and $\Emax$ are displayed in blue, red and brown, respectively.

In all calculations, the pW values are larger than the corresponding pE values. Due to the quadratic convergence of the correlation energy, pE values were found between roughly 2--4 units smaller than pW values. 

Clearly, all CI results show identical pE and pW values for the different operator sets $\ELow$, $\Emin$ and $\Emax$. Since all operator sets span the  same linear space, this result was expected for a linear theory such as CI and acts as a proof of concept for the equality of the spanned linear spaces. 

For CC, the results show significant differences in both pE and pW values for the different operator sets. Overall, the best results are obtained for the $\ELow$ and $\Emin$ operator sets. Most of the results for the operator set $\Emax$, lead to smaller pE and pW values compared to the $\ELow$ and $\Emin$ operator sets. In some cases the differences are significant. The CCSDT($\Emax$) calculation in subfigure \ref{N_4_3_CC.fig} led to a 2.14 units lower pW as well as a 1.42 units lower pE compared to CCSDT($\Emin$) and CCSDT($\ELow$). In absolute (relative) values this means an error in the correlation energy of ca. $0.12\,$mH (ca. $0.32\%$). While the absolute numbers seem to be negligible for this particular example, a relative error of $0.32\%$ in the correlation energy may be of critical importance when approaching chemical accuracy in larger molecular systems. 

The two operator sets $\Emin$ and $\ELow$ show identical or very similar pE and pW values. Since both $\Emin$ and $\ELow$ possess a minimal amount of spectators (while still being composed of different operators), it may be assumed that only the amount of spectators is responsible for the observed accuracy loss. 

In direct comparison to CI, the pE and pW results of CC converge faster or as fast as CI towards the FCI limit. In no case, a substantially slower CC convergence was observed. An increased number of spectators (as e.g. used in the CC($\Emax$) calculations) slowed down the CC convergence of pE and pW towards the FCI limit but did never lead to a slower convergence than observed with CI. For the CI and CC calculations of $^3$P carbon shown in subfigures \ref{C_3_2_CI.fig} and \ref{C_3_2_CC.fig}, respectively, this is clearly recognizable. CI pW values converge with 1.81, 3.88, 4.57, 8.47 and 9.78\,units towards the FCI limit while CC($\Emax$) pW values converge only slightly faster with 1.81, 3.95, 5.16, 8.83 and 10.80\,units to the FCI limit when compared to e.g. CC($\ELow$) with 1.81, 3.99, 6.72, 10.13 and 13.00\,units.

\begin{center}
\begin{figure}[h]
\includegraphics[scale=1]{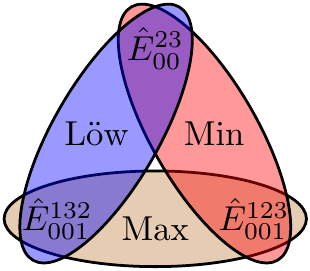}
\caption{\label{ThreeOperators.fig}\raggedright Affiliation diagram of the three operators $\hat{E}_{00}^{23}$, $\hat{E}_{001}^{123}$ and $\hat{E}_{001}^{132}$ to the $\ELow$, $\Emin$ and $\Emax$ operator sets.}
\end{figure}
\end{center}

The reason for this unfavorable behavior of the SASC-CC($\Emax$) calculations can be found in the non-linear product terms $\hat{T}^2$, $\hat{T}^3$, etc. For an increased amount of spectators, an increased amount of these product terms vanishes. Ultimately, for the worst case $\hat{T}^n = 0 \quad\forall_{n>1}$ this leads to a completely linear CC wave function ansatz analogously to CI. Consider e.g. the spatial configuration $\ket{1}\ket{2}\ket{3}$ composed of spatial orbitals 1, 2 and 3 reachable from the doublet reference CSF $\ket{0\overline{0}1}$ via a double substitution. Here we denote alpha and beta spin states via non-overlined and overlined indices. The spin degeneracy for 3 open shells and $S=S_z=\frac{1}{2}$ requires 
\begin{equation*}
f\left(3,\frac{1}{2}\right) = \binom{3}{1} - \binom{3}{0} = 2\,
\end{equation*}
operators to span the complete spin space. In total, the three operators 
\begin{align*}
\hat{E}_{00}^{23}  \ket{0\overline{0}1} &= \ket{12\overline{3}} - \ket{1\overline{2}3}\, \\
\hat{E}_{001}^{123}\ket{0\overline{0}1} &= \ket{1\overline{2}3} - \ket{\overline{1}23}\text{ and} \\ 
\hat{E}_{001}^{132}\ket{0\overline{0}1} &= \ket{\overline{1}23} - \ket{12\overline{3}}
\end{align*}

appear in the operator sets $\ELow$, $\Emin$ and $\Emax$ as outlined in Figure \ref{ThreeOperators.fig}. 

The operator sets $\ELow$ and $\Emin$ are composed of the minimal spatial substitution $\hat{E}_{00}^{23}$ and the second spatial substitution $\hat{E}_{001}^{132}$ ($\ELow$) or $\hat{E}_{001}^{123}$ ($\Emin$) incorporating the spectating index $1$. In contrast, the operator set $\Emax$ does not include the minimal spatial substitution $\hat{E}_{00}^{23}$ but only the two spectators. Any product with a single substitution $\hat{E}_1^a$ from the right, leads to 
\begin{align*}
\hat{E}_{00}^{23}\hat{E}_1^a\ket{0\overline{0}1} &= (\delta_{a3} - 1)\ket{\overline{2}3a} + (1 - \delta_{a2})\ket{2\overline{3}a}\\
\hat{E}_{001}^{123}\hat{E}_1^a\ket{0\overline{0}1} &= 0 \\
\hat{E}_{001}^{132}\hat{E}_1^a\ket{0\overline{0}1} &= 0\,
\end{align*}
such that products $(\ldots )\hat{E}_1^a$ completely vanish for the operator set $\Emax$. In case of the operator sets $\ELow$ and $\Emin$ this is not entirely the case since the product $\hat{E}_{00}^{23}\hat{E}_1^a$ does not vanish for arbitrary indices $a\in\mathbb{V}$.

\subsection{Invariance w.r.t. Orthogonalization for Spin Complete Operator Sets}
\label{(B)}

As an additional validation of the present implementation we checked for invariance of the results w.r.t. the
orthogonality of the spin operators for CI and CC approaches.
All spin-complete orthogonal CI/CC results are identical to the spin-complete non-orthogonal CI/CC results.
Therefore, an analogous discussion to the influence of different amounts of spectators in subsection \ref{(A)} 
may be applied to the orthogonalized operator sets $\hat{\mathbb{E}}_{\text{L\"ow}}'$, $\hat{\mathbb{E}}_{\text{min}}'$ 
and $\hat{\mathbb{E}}_{\text{max}}'$.

Results of all conducted CI as well as CC calculations using the orthonormalized operator sets $\ELow'$, $\Emin'$ and $\Emax'$ 
can be found in the supporting information. 

\subsection{Analysing the Influence of Spin Completeness}
\label{(C)}

As explained in our previous work\cite{Herrmann2020a}, $\hat{E}$ operator spectators are necessary to reach spin-complete operator sets for open-shell spin-adapted CC calculations. Without the minimal required amount of spectators, a significant spin incompleteness error can arise. To systematically investigate this spin incompleteness effect, we defined three different cluster/CI operator truncation sublevels $(0)$, $(1)$ and $(2)$. Each truncation sublevel may be applied to each operator set $\hat{\mathbb{E}}_{M}$ for each spatial substitution order $\nu$. Here,

\begin{itemize}
\item[$(0)$] denotes the unaltered fully spin-adapted and spin-complete operator set $\hat{\mathbb{E}}_{M}^{(0)} = \hat{\mathbb{E}}_{M}$,
\item[$(1)$] denotes a spin-adapted but possibly spin-incomplete operator set $\hat{\mathbb{E}}_{\text{M}}^{(1)}$, where only spectators of nominal tensor rank $\mu$ with $\mu \le \nu$ are included and 
\item[$(2)$] denotes a spin-adapted but spin-incomplete operator set including no spectators at all.  
\end{itemize}

In Table \ref{SublevelExample.tab}, the affiliation of different single and double spatial substitution operators from the operator set $\Emin$ to the sublevels (0), (1) and (2) in the spatial singles (S) as well as singles and doubles (SD) truncation is shown. 

\begin{center}
\begin{table}[H]
\centering
\caption{\label{SublevelExample.tab} Truncation sublevel affiliation of different spatial substitution operator prototypes of spatial substitution order 1 and 2.}
\begin{tabular}{c|c|c||c|c|c||c|c|c}
\centering
& & & \multicolumn{3}{c||}{\textbf{S}} & \multicolumn{3}{c}{\textbf{SD}} \\
\textbf{Order} $\nu$ & \textbf{Rank} & \textbf{Operators} &  (0) & (1) & (2) & (0) & (1) & (2)\\ \hline
\multirow{2}{*}{1} & 1 & $\hat{E}_i^a$, $\hat{E}_i^v$, $\hat{E}_v^a$ & \checkmark & \checkmark & \checkmark & \checkmark & \checkmark & \checkmark \\ \cline{2-9} 
  & 2 & $\hat{E}_{iv}^{va}$ & \checkmark &  &  & \checkmark & \checkmark &  \\ \hline \hline
 & 2 & $\hat{E}_{ii}^{aa}$, $\hat{E}_{ii}^{ab}$, $\ldots$ &  &  &  & \checkmark & \checkmark & \checkmark \\ \cline{2-9} 
2 & 3 & $\hat{E}_{ijv}^{vab}$, $\hat{E}_{ijv}^{avb}$, $\ldots$ &  &  &  & \checkmark &  &  \\ \cline{2-9} 
 & 4 & $\hat{E}_{ijvw}^{vwab}$ &  &  &  & \checkmark &  &
\end{tabular}
\end{table}
\end{center}

For the SD truncation, sublevel (0) includes all spectators -- also operators of tensor ranks 3 and 4 for doubles --, sublevel (1) readily stops at tensor rank 2 operators but includes all rank 2 spectators (of spatial substitution order $\nu = 1$) and sublevel (2) includes no spectators at all. Clearly, sublevels (1) and (2) span an increasingly insufficient spin space. While sublevel (1) retains at least some spin-complete parts, sublevel (2) is spin-incomplete per se. Furthermore, sublevels (0) and (1) are identical for the full cluster/CI operator, i.e. converge towards the exact FCI limit.

Despite their potential incompleteness, all defined operator subsets are made up of spatial substitutions $\hat{E}$ only and therefore produce pure spin states. This may be checked by calculating the spin projection error as defined in subsection \ref{SpinProjectionError}. In Table \ref{SpinError.tab}, spin projection errors $\epsilon$ for sublevels (0), (1) and (2) employed in a CC calculation are compared to spin projection errors of standard spinorbital CC for the atomic $^3$P carbon. Spin projection errors for all conducted calculations can be found in the supporting information.

\begin{center}
\begin{table}[H]
\centering
\caption{\label{SpinError.tab}Summarized spin projection errors for spinorbital and spin-adapted non-orthogonal CC calculations employing the cluster operator truncation sublevels (0), (1) and (2) for the operator set $\hat{\mathbb{E}}_{\text{L\"ow}}$ of atomic $^3$P carbon in the 6-31G\cite{Hehre1972} basis set.}
\begin{tabular}{c|c|c|c|c}
\centering
& \textbf{Spinorbital} & \multicolumn{3}{c}{\textbf{Spin-Adapted CC}} \\
\textbf{Truncation} & \textbf{CC} & \textbf{(0)} & \textbf{(1)} & \textbf{(2)} \\ \hline
S      & $2.07\cdot 10^{-2\phantom{6}}$ & $<10^{-17}$ & $<10^{-17}$ & $<10^{-17}$ \\
SD     & $1.79\cdot 10^{-3\phantom{6}}$ & \textquotedbl & \textquotedbl & \textquotedbl \\
SDT    & $8.26\cdot 10^{-5\phantom{6}}$ & \textquotedbl & \textquotedbl & \textquotedbl \\
SDTQ   & $1.75\cdot 10^{-6\phantom{6}}$ & \textquotedbl & \textquotedbl & \textquotedbl \\
SDTQ5  & $1.96\cdot 10^{-8\phantom{6}}$ & \textquotedbl & \textquotedbl & \textquotedbl \\
SDTQ56 & $3.46\cdot 10^{-16}$ & \textquotedbl & \textquotedbl & \textquotedbl
\end{tabular}
\end{table}
\end{center}

All calculations including sublevels (0), (1) and (2) lead to spin projection errors of zero (within the double precision limit) proving their spin state purity. In contrast, spinorbital CC calculations lead to spin projection errors in the order of $10^{-2}$ to $10^{-8}$ for truncation S to SDTQ5. For an untruncated cluster operator (SDTQ56, the FCI limit), a spin projection error of zero is obtained. 

To investigate the effect of the different sublevels (0), (1) and (2) on the wave function as well as the amount of recovered correlation energy, pW and pE values were calculated according to equations \ref{pW.eq} and \ref{pE.eq}. Results of all conducted calculations (Li to N) can be found in the supporting information.

\begin{center}
\begin{figure}[H]
\centering 
\subfloat[\label{N_4_3_III_OCC.fig}$^4$S nitrogen (Orth. CC)]{\includegraphics[scale=.95]{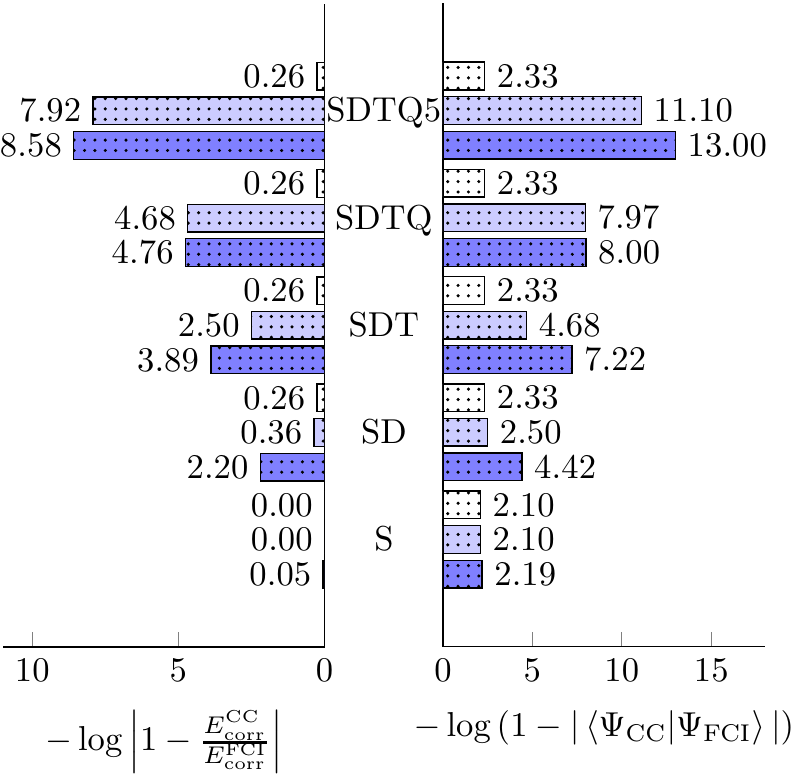}}
\quad
\subfloat[\label{N_4_3_III.fig}$^4$S nitrogen (Non-Orth. CC)]{\includegraphics[scale=.95]{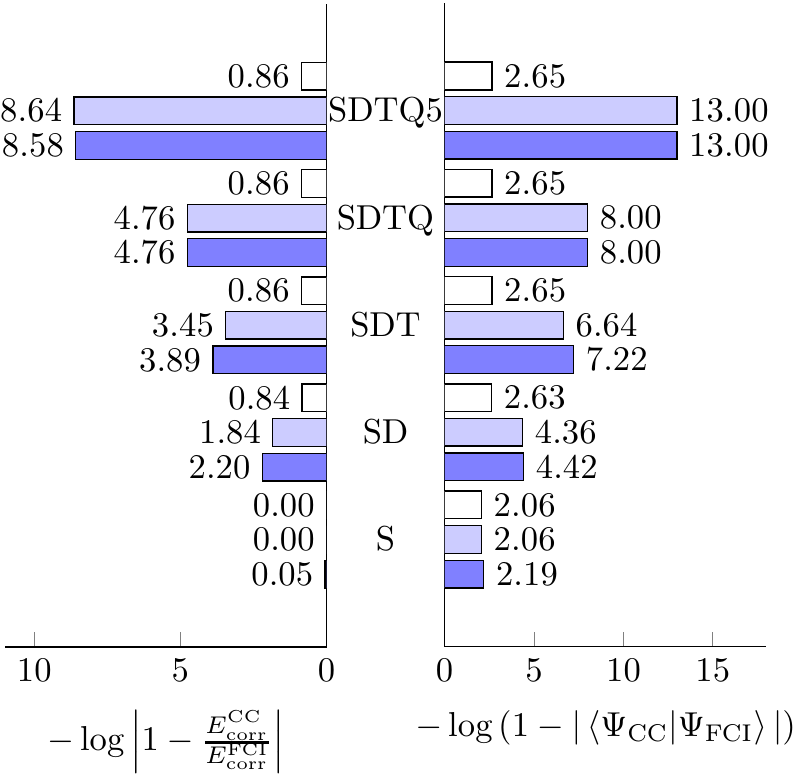}}
\\
\subfloat[\label{N_6_5_III_OCC.fig}$^6$P nitrogen (Orth. CC)]{\includegraphics[scale=.95]{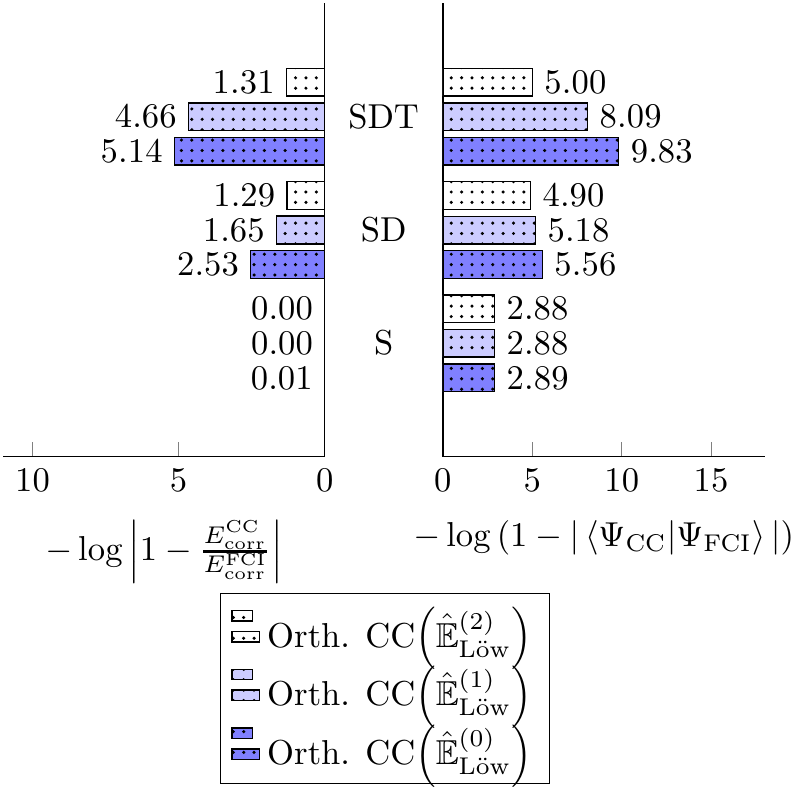}}
\quad
\subfloat[\label{N_6_5_III.fig}$^6$P nitrogen (Non-Orth. CC)]{\includegraphics[scale=.95]{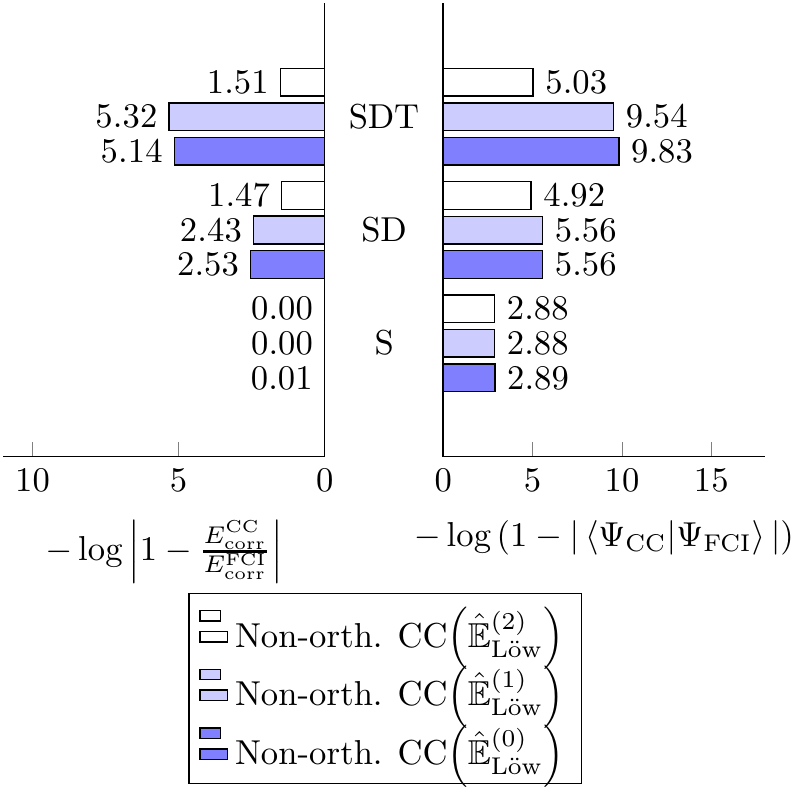}}
\caption{\label{N_III.fig}Summarized pE and pW values for spin-adapted and spin-complete orthogonal (left) as well as non-orthogonal (right) CC calculations for atomic $^4$S and $^6$P nitrogen in the 6-31G\cite{Hehre1972} basis set employing operator sublevels (0), (1) and (2). In all calculations, the $\ELow$ operator set was used.}
\end{figure}
\end{center}

In Figure \ref{N_III.fig}, the obtained results for the $^4$S and $^6$P high spin states of atomic nitrogen are shown.
The shown results were obtained from orthogonal (left) and non-orthogonal (right) CC calculations employing the operator set $\hat{\mathbb{E}}_{\text{L\"ow}}$. As in the previous subsections, bar charts advancing to the right show pW values while
 bar charts advancing to the left show pE values. To distinguish among the three truncation sublevels (0), (1) and (2), blue colors with decreasing opacity were used. Here, dark blue resembles truncation sublevel (0), light blue sublevel (1) and white sublevel (2). 
 
 Clearly, truncation sublevel (2) leads to very inaccurate results compared to sublevels (0) and (1). For the shown example (neglecting singles), pE values are found between 1.06 -- 8.32\,units smaller than sublevels (0) and (1) while pW values are found between 0.64 -- 10.76\,units smaller than sublevels (0) and (1). The comparison of sublevels (0) or (1) and sublevel (2), is however not on

\onecolumngrid
\twocolumngrid

the same footing since even for a full spatial substitution rank $\nu = n$, sublevel (2) can never reach the FCI limit while sublevels (0) and (1) readily converge to the same FCI solution. 

Comparing sublevels (0) and (1), only minor differences are recognizable in the non-orthogonal case. There are however certain exceptions like e.g. $^4$S nitrogen CCSD. Here, pE values of $2.20$ and $1.84$ are obtained, which indicate 99.37\% and 98.42\% of recovered correlation energy (w.r.t. FCI), respectively. This error of roughly 1\% in the correlation energy might be of importance when approaching chemical accuracy. In most of the conducted non-orthogonal calculations however, the difference between sublevels (0) and (1) seems to be negligible. This effect is of particular interest since the high rank spectators in sublevel (0), which are neglected in sublevel (1), are expected to be the computationally most expensive terms of the CC equations.
Since only small atomic calculations were conducted in this work, this effect might not necessarily carry on to larger molecular systems. Here, the difference between these sublevels might be of greater importance. This will be subject to further investigations.

In the orthogonal case, the results of sublevel (0) (as discussed in subsection \ref{(B)}) are identical to the non-orthogonal results. For sublevels (1) and (2), spectators of different tensor ranks are neglected. In the orthogonal cluster operator, which is composed of linear combinations of spatial substitutions, all orthogonal substitutions including at least one neglected spectator are completely ignored. 
Therefore, sublevels (1) and (2) may lead to fewer amplitudes in the orthogonal case compared to the non-orthogonal case
resulting in smaller pE and pW values in the orthogonal sublevels (1) and (2) compared to the non-orthogonal case.

\section{Conclusion}

In this work, atomic test calculations of spin-adapted CI and CC were investigated in terms of 
\begin{itemize}
\item[(a)] the wave function quality (through means of the overlap to the FCI wave function) and 
\item[(b)] the amount of recovered correlation energy (w.r.t. FCI). 
\end{itemize}

Both (a) and (b) were compared for CI and CC calculations using different sets of spatial substitution operators composed of different amounts of spectators as well as orthogonal and non-orthogonal substitutions. 

In section \ref{(A)}, the influence of different amounts of spectators in spatial substitution operator sets spanning identical linear spaces was investigated. In total, three different but same-space-spanning operator sets $\hat{\mathbb{E}}_{\text{L\"ow}}$, $\hat{\mathbb{E}}_{\text{min}}$ and $\hat{\mathbb{E}}_{\text{max}}$ were compared. While $\hat{\mathbb{E}}_{\text{min}}$ and $\hat{\mathbb{E}}_{\text{max}}$ were composed of a minimal and maximal amount of spectators, the set $\hat{\mathbb{E}}_{\text{L\"ow}}$ was generated by an automated operator generation approach explicitly discussed in our previous work\cite{Herrmann2020a}. As expected, all CI calculations show identical results for all operator sets, 
while CC (due to its non-linear wave function ansatz) shows differences in both the FCI overlap and the amount of recovered correlation energy.
In the present study, the best results were found for the operator sets $\hat{\mathbb{E}}_{\text{L\"ow}}$ and $\hat{\mathbb{E}}_{\text{min}}$, which were almost identical in most scenarios. Operator set $\hat{\mathbb{E}}_{\text{max}}$ led to significantly flawed CC results with errors in the correlation energy of up to 0.32\% (w.r.t. FCI).

In section \ref{(B)}, the CC results for sets of orthogonal and non-orthogonal spatial substitution operators were compared. As explicitly discussed in subsection \ref{Orthogonality}, cluster operators linked by an invertible linear transformation should lead to identical results. As expected, identical results for spin-complete orthogonal and non-orthogonal CC were obtained. Future open-shell spin-adapted CC developments may therefore benefit from the usage of the simpler non-orthogonal operators \cite{Herrmann2020a} over the more complicated orthogonalized form without any accuracy gain or loss.

Finally, in section \ref{(C)} the effect of spin incompleteness on the correlation energy and the wave function quality was investigated. Three truncation sublevels (0), (1) and (2) for an increasingly insufficient spin space were defined for each spatial substitution truncation level. Here, sublevel (0) denotes the fully spin-complete cluster operator, sublevel (1) neglects operators with nominal tensor ranks higher than the applied spatial truncation while sublevel (2) denotes a strongly spin-incomplete cluster operator including no spectators at all. The results for sublevels (0) and (1) were found to be qualitatively identical for non-orthogonal cluster operators with quantitative agreement in most cases. The results for sublevel (2) however, showed large errors of several orders of magnitude in both the recovered correlation energy and the wave function emphasizing the importance of spin completeness in CI and CC calculations. 

\begin{appendix}
\section{Overlap Maximization}
\label{appendix}
Consider a set $\left\{\ket{\Phi_i}\right\}$ for $i=1,\ldots m$ of degenerate eigenfunctions of the Hamiltonian with 
\begin{equation}
\hat{\mathcal{H}}\ket{\Phi_i} = E\ket{\Phi_i} \quad\forall_i\,,
\end{equation}
for which 
\begin{equation}
\braket{\Phi_i | \Phi_j} = \delta_{ij}
\end{equation}
holds. Due to the linearity of the Hamiltonian it is clear that arbitrary linear combinations $\ket{\Psi}$ with 
\begin{equation}
\ket{\Psi} = \sum_{i=1}^m c_i \ket{\Phi_i} 
\end{equation}
are also eigenfunctions of $\hat{\mathcal{H}}$ with the same eigenvalue $E$ -- hence the rotational freedom of the eigenspace. For a given CC or CI wave function $\ket{\tilde{\Psi}}$, the overlap to $\ket{\Psi}$ may be maximized under the constraint of the latter remaining normalized. An appropriate Lagrangian $\mathcal{L}$ takes the form 
\begin{equation}
\mathcal{L}(c_1, \ldots c_m, \lambda) = \sum_{i=1}^m c_i \braket{\tilde{\Psi} | \Phi_i} + \lambda\left(\sum_{i=1}^m |c_i|^2 - 1\right)\,.
\end{equation}
Assuming real coefficients $\left\{c_i\right\}$, the partial derivatives with respect to $c_j$ and $\lambda$ are given by 
\begin{align}
\label{delci.eq}
\frac{\partial \mathcal{L}(c_1,\ldots c_m,\lambda)}{\partial c_j} &= \braket{\tilde{\Psi} | \Phi_j} + 2\lambda c_j \stackrel{!}{=} 0 \quad \forall_{j\in\{1\ldots m\}} \\ 
\frac{\partial \mathcal{L}(c_1,\ldots c_m,\lambda)}{\partial \lambda} &= \sum_{i=1}^m c_i^2 - 1 \stackrel{!}{=} 0\,,
\label{dellambda.eq}
\end{align}
respectively. From equation (\ref{delci.eq}) for $j=1\ldots m$ it is 
\begin{equation}
\label{cj.eq}
c_j = - \frac{\braket{\tilde{\Psi} | \Phi_j}}{2\lambda}\,.
\end{equation}
Insertion into (\ref{dellambda.eq}) then leads to 
\begin{equation}
\label{lambda.eq}
\lambda = \pm \sqrt{\sum_{i=1}^m \frac{\braket{\tilde{\Psi} | \Phi_i}^2}{4}}\,,
\end{equation}
such that the maximized (minimized) overlap $S$ is given by
\begin{equation}
S = \sum_{i = 1}^m c_i \braket{\tilde{\Psi} | \Phi_i} = \frac{\sum_{i=1}^m \braket{\tilde{\Psi} | \Phi_i}^2}{-2\lambda}
\end{equation}
for a negative (positive) choice of $\lambda$.
\end{appendix}

\end{document}